\documentclass{article}

\usepackage{arxiv}

\usepackage[utf8]{inputenc} 
\usepackage[T1]{fontenc}    
\usepackage{hyperref}       
\usepackage{url}            
\usepackage{booktabs}       
\usepackage{amsfonts}       
\usepackage{nicefrac}       
\usepackage{microtype}      
\usepackage{lipsum}		
\usepackage{graphicx}
\usepackage[square,numbers]{natbib}
\usepackage{doi}

\usepackage[export]{adjustbox}
\usepackage{tablefootnote}
\usepackage{xcolor}
\usepackage{graphicx}
\usepackage{cuted}
\usepackage{float}
\usepackage{graphicx}
\usepackage{caption}
\usepackage{comment}
\usepackage{amsmath}
\definecolor{color}{RGB}{25,25,112}
\definecolor{negro}{RGB}{0,0,0}
\definecolor{colorurl}{RGB}{25,25,112}
\usepackage{orcidlink}
\usepackage{amssymb}
\usepackage{tikz}
\usepackage{tikz-3dplot}
\usetikzlibrary{patterns}

\title{Exciton Coulomb correlations in axially symmetric II-VI nanocrystals}

\author{ \href{https://orcid.org/0009-0006-1840-649X}{\includegraphics[scale=0.06]{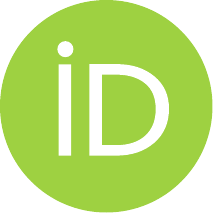}\hspace{1mm} Carlos A. Boh{\'o}rquez-Quincos} \href{https://orcid.org/0000-0000-0000-0000}{\includegraphics[scale=0.06]{orcid.pdf}\hspace{1mm} Juli\'an G. Cordero-Niño} \href{https://orcid.org/0000-0003-1575-6583}{\includegraphics[scale=0.06]{orcid.pdf}\hspace{1mm} Hanz Y. Ram\'irez-G\'omez} \thanks{hanz.ramirez@uptc.edu.co} \\
	Grupo de F\'isica Te\'orica y Computacional \& Grupo QUCIT, \\
	Escuela de F\'isica, Universidad Pedag\'ogica y Tecnol\'ogica de Colombia (UPTC),\\
	Tunja 150003, Boyac\'a, Colombia. \\
}

\hypersetup{
	pdftitle={Exciton Coulomb correlations in axially symmetric QDs},
	pdfsubject={q-bio.NC, q-bio.QM},
	pdfauthor={C. A. Boh{\'o}rquez-Quincos, H. Y. Ram\'irez-G\'omez},
	pdfkeywords={Excitons, Coulomb Correlations, Semiconductor Nanoparticles},
}

\begin{document}
	\maketitle
	
\begin{abstract}
Because of their conveniently tunable optoelectronic properties, semiconductor nanocrystals have become established components for  new devices and emergent technologies, in a broad range of applications which include agriculture, medicine, energy harvesting and quantum information. 

In this work we study the dielectric and quantum confinement effects on excitons confined in single axially symmetric quantum dots, and systematically calculate by means of the configuration interaction method the corresponding Coulomb correlations in the cases of CdSe, CdTe and ZnTe nanocrystals. 

We show that effects from the dielectric confinement, although one order of magnitude bigger than those from the quantum confinement, are negligible in the exciton energies because of mutual cancellation of the electron and hole contributions. 

Regarding the electrostatic interaction, we found for all analyzed cases that the exciton binding energy is below 2\% of the non-interacting electron-hole pair energy. Furthermore, our calculations reveal that the in general the first-order-perturbation correction accounts for 70 \% to 90 \% of the binding energy. 

These results provide useful information on the magnitude of the confinement effects in II-VI semiconductor nanocrystals, suggesting that a simple perturbative approach may be accurate enough for including the electrostatic interaction in most cases. 
\end{abstract}


\section{Introduction}
Throughout the 20th century, immense development was achieved in the technological field with the invention of the transistor, a semiconductor electronic device that operates by the junction of two N-type and P-type materials, allowing the flow of a signal between two of its contacts (from collector to emitter) depending on the input in its third contact (base) \citep{transistor}. This behavior, along with the inherent nature of semiconductor materials, gradually led to a decrease in transistor sizes over the last few decades, reaching the nanometer scale, in which quantum behaviors should be taken into consideration.

Since the discovery of quantum dots (QDs), the techniques for synthesizing them have been widely improved and diversified, allowing us to count on QDs with different characteristics and made of different materials. These include Core-type QDs, core-shell QDs, and hollow QDs \cite{conito,vaso,panela}.

To advance our understanding of Coulomb interaction in QDs, we begin by addressing an article published in 1990 by Hu et al., which presented a mathematical description of quantum confinement based on the effective mass model \cite{hu}. They also calculated the effects of Coulomb interaction in an ideal quantum confinement, which resulted in an increase in the biexcitonic energy as the QD size decreases.

In 1996, Shu-Shen Li et al. used the effective mass model to study strained InAs/GaAs coupled QDs, investigating excitonic states, energy transitions with respect to size, and optical absorption spectra \cite{Li}. These results were consistent with those presented by Wang et al. in 1994 \cite{wang}.

In 1998, J.M. Ferreyra presented a work similar to that of Shu et al. in 1996, considering aspects such as the finite height of confinement barriers and the difference in effective masses of electrons and holes inside and outside the QD. They obtained improved theoretical results for the excitonic exchange energies \cite{ferre}.

In 1998, an article by M. Rontani et al. focused on studying the energy spectra of three-dimensional and two-dimensional semiconductor QDs using two approximation schemes: the Hartree-Fock model and the Hubbard model \cite{hubb}. It was shown that both models provided suitable descriptions for Coulomb correlation, but the results were more accurate for the 3D case \cite{ronta}.
En 2004, Baer et al. constructed the Hamiltonian for a cylindrically symmetric quantum dot system, where the correlation terms can be expressed as matrix elements. Due to the complexity of the system, it was reduced based on the importance of the different types of terms \cite{baer}. Later, in 2006, Baer et al., building on their previous work, conducted an analysis of the optical properties of quantum dots based on nitrides. These nanocrystals exhibit skew excitons \cite{baer2}. Once again, the use of Coulomb matrix elements was essential, as they served as input parameters for calculating configuration interaction effects and describing optical properties.

In 2010, Chen et al. in their paper "Engineered spin-state transitions of two interacting electrons in semiconductor nanowire quantum dots" derived an analytical expression for Coulomb matrix elements in axially symmetric quantum dots using a parabolic model. This analytical expression was further implemented and improved for systems including an external magnetic field by Ramirez H. et al. in 2015, in a computational search for second harmonic generation in QDs, demonstrating the effectiveness and low computational cost of matrix elements in describing Coulomb effects and correlations in a quantum dot \cite{chen, Profe}. It is important to note that these studies have also been extended to 2D nanostructures \cite{xu, 2dim}, and in some specific cases, such as CdTe, Coulomb interactions do not contribute significantly due to its large Bohr radius \cite{mez}.

This historical review highlights the evolution of models considering Coulomb effects in quantum dots and how they are relevant in various research fields, such as harmonic generation, polarization energy manipulation, generation of entangled photons, electron spin schemes for qubit modeling, and single photon sources \cite{chen, Hanzp, harmonic, spin, springer}. It is worth mentioning that while several previous works provide methods to quantify interactions, to the best of our knowledge there is no comprehensive study that systematically compares and evaluates these interactions for II-VI ellipsoidal quantum dots of different volumes and aspect ratios.

This document is composed by three main sections. In the first section, the electron-hole model in an axially symmetric single quantum dot is presented, together with its corresponding solution. The second chapter discusses the theoretical and computational results derived from this model. In the third chapter, the approach is extended to the study of a vertically coupled double quantum dot. Finally, the paper ends with a summary of the conclusions obtained throughout the work, followed by a section of appendices containing all the necessary resources for the development of the research.

\section{Theorical model}
\label{sec:model}
Our system consists of an electron-hole pair confined in an axially symmetric quantum dots, with Hamiltonian 

\begin{align} \label{H}
	\hat{H}=\hat{H}_e+\hat{H}_h+\hat{H}_{eh}+\frac{1}{2}\hat{H}^{self}_{e}+\frac{1}{2}\hat{H}^{self}_{h},
\end{align}

\noindent where $\hat{H}_e$  and $\hat{H}_h$ represent the kinematic and quantum confinement part in the Hamiltonian of the electron and hole, respectively. $\hat{H}_{eh}$ accounts for the Coulomb effects between hole and electron. These terms are taken as three-dimensional harmonic oscillator Hamiltonians because this approximation provides a very good depiction of the confinement of the charged particles in a quantum dot, the last two terms $\hat{H}^{self}_{e}$ and $\hat{H}^{self}_{h}$ represents the effect by a point charge (electron-hole) due to dielectric confinement in a the spheroidal shape of quantum dot.

The eigenvalues to the hamiltonian can be evaluated individually, as:
\begin{align}\label{TotalEnergy}
	E_e+E_h+E_{e-h}+E_{e-e}^{self}+ E_{h-h}^{self}+E_{e-h}^{self}+E_{h-e}^{self}	,
\end{align}
In this document we will see how to find the first 3 energy terms, the energy elements labels with "self" represents  self-energy due to dielectric confinement and to calculate the value of these terms we must multiply the charge of the particle by the potential that it generates, which can be found by means of Method of Images \cite{jackson2021}. The mathematical form for the "self" elements are, 
\begin{align}
	&E_{e-e}^{self}=\frac{q^2}{2}\sum C_n Fn,\\&E_{h-h}^{self}=\frac{q^2}{2}\sum C'_n F'n,
\end{align}
where $F'n$ and $Fn$ are polynomials derived from solving the Laplace equation in addiction $C'_n$ and $C_n$ are coefficient, which can be founded from the boundary conditions, the two terms are positive, since both the hole and the electron generate a potential of the same sign and electrically they are the same $E_e^{self}=E_h^{self}$. 
By another hand  
\begin{align}
	&E_{e-h}^{self}=\frac{-q_e*q_h}{2}\sum C_n Fn,\\  \text{and} &\\
	&E_{h-e}^{self}=\frac{-q^2}{2}\sum C'_n F'n,
\end{align}
acounts the effects  between the potential generate for a carrier charge and the potential due to the image charge generated by polarization, we get four energy contribution with the same magnitude but with different sign, therefore, we can conclude that the "self" energy terms in the equation (\ref{TotalEnergy}) do not contribute because they cancel each other out and the dielectric confinement may be neglected.

\begin{figure}
\centering
\adjustbox{trim=0 1.2cm 0 0 , clip}
{\includegraphics[width=0.5\textwidth]{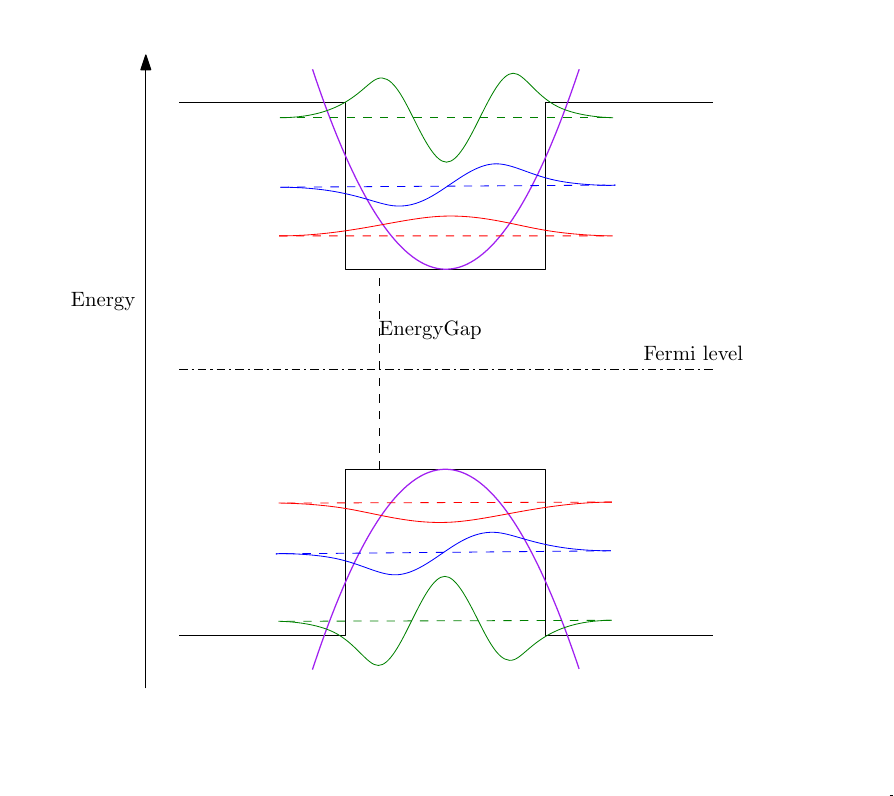}}
\caption{Eigenstates in  harmonic oscillator approximation for a Single QD.}
\label{fig:lightcurve}
\end{figure}

\subsection{Dielectric confinement}
\label{sec:Dielectric}
The matrix elements for the dielectric confinement for an electron (hole) are 

\begin{align}\label{self}
	\hat{H}_{e/h}^{self}=\frac{1}{2}\langle \Psi_{m\prime_{e/h},n\prime_{e/h},s\prime_{e/h}}|(\mp q) \phi(\vec{r}\,)|\Psi_{m_{e/h},n_{e/h},s_{e/h}}\vec{r}\,)\rangle , 
\end{align}

\noindent where $\phi(\vec{r})$ is the electrostatic potential, for a specific confinement geometric, for this word we take the three shapes for a single Quantum dots as shown in Figure (\ref{fig:Asp}), and for explicit calculations in $ \hat{H}_{e/h}^{self}$ we take the eigenstates of ground state to make a first approximation.
\subsubsection{Dielectric confinement for a sphere}
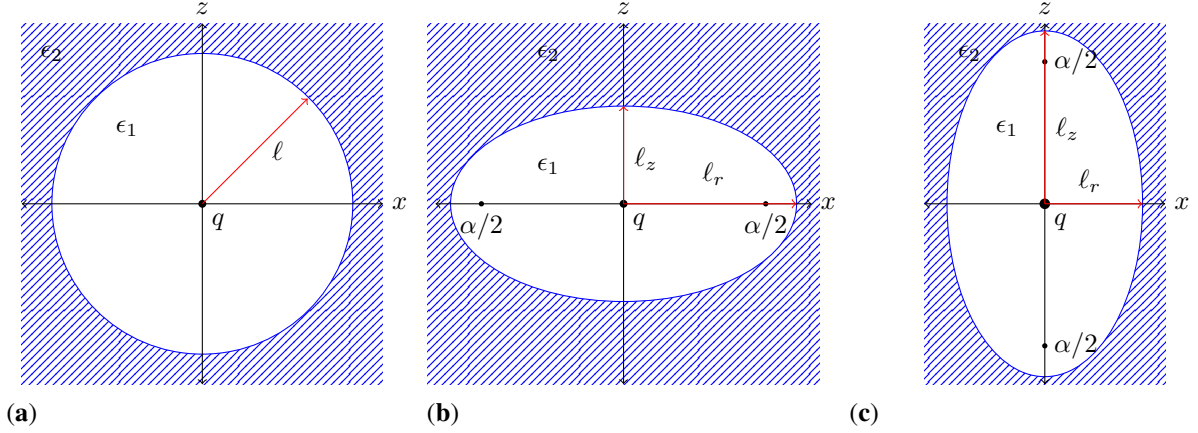
\begin{figure*}
	\centering 
	\begin{tikzpicture}
		\begin{scope}[shift={(-5.6,0)}] 
			\begin{scope}\clip (-2.4,-2.4) rectangle (2.4,2.4);            \fill[pattern=north east lines, pattern color=blue] (-3,-3) rectangle (3,3); 
			\end{scope}        
		\draw[draw=blue,fill=white] (0,0) circle (2cm); 
		\fill (0,0) circle (1.5pt) node[below right] {$q$};     
		\node at (-1, 1) {$\epsilon_1$}; 
		\node at (-2, 2) {$\epsilon_2$}; 
		\draw[<->] (-2.4,0) -- (2.4,0) node[right] {$x$}; 
		\draw[<->] (0,-2.4) -- (0,2.4) node[above] {$z$}; 
		\draw[->,draw=red] (0,0) -- (1.4,1.4);
		\node at (1, 0.7) {$\ell$};
		\node at (-2.4,-2.8) {($\textbf{a}$)};
	\end{scope}
	\begin{scope}[shift={(0,0)}] 
		\begin{scope}
			\clip (-2.6,-2.4) rectangle (2.6,2.4);
			\fill[pattern=north east lines, pattern color=blue] (-3,-3) rectangle (3,3);
		\end{scope}
		\draw[draw=blue,fill=white] (0,0) ellipse (2.3cm and 1.3cm); 
		\fill (0,0) circle (1.5pt) node[below right] {$q$};
		\fill (-1.89,0) circle (1pt) node[below] {$\alpha/2$};        				  \fill (1.89,0) circle (1pt) node[below ] {$\alpha/2$};
		\node at (-1, 0.5) {$\epsilon_1$}; 
		\node at (-1, 2) {$\epsilon_2$}; 
		\draw[<->] (-2.5,0) -- (2.5,0) node[right] {$x$}; 
		
		\draw[<->] (0,-2.4) -- (0,2.4) node[above] {$z$}; 
		
		\draw[->,draw=red] (0,0) -- (0,1.3);
		\node at (0.3, 0.6) {$\ell_z$};
		\draw[->,draw=red] (0,0) -- (2.3,0);
		
		\node at (1.2,0.4) {$\ell_r$};
		\node at (-2.4,-2.8) {($\textbf{b}$)};
	\end{scope}
	\begin{scope}
		[shift={(5.6,0)}] 
		\begin{scope}
			\clip (-1.6,-2.4) rectangle (1.6,2.4);            			  	 			\fill[pattern=north east lines, pattern color=blue] (-1.6,-3) rectangle (1.6,3);
		\end{scope}
		\draw[draw=blue,fill=white] (0,0) ellipse (1.3cm and 2.3cm); 
		\fill (0,0) circle (2pt) node[below right] {$q$}; 
		\fill (0,-1.89) circle (1pt) node[right] {$\alpha/2$};         			\fill (0,1.89) circle (1pt) node[right ] {$\alpha/2$};       				\node at (-0.5, 1) {$\epsilon_1$}; 
		\node at (-1, 2) {$\epsilon_2$}; 
		\draw[<->] (-1.6,0) -- (1.6,0) node[right] {$x$}; 
		\draw[<->] (0,-2.4) -- (0,2.4) node[above] {$z$}; 
		\draw[->,draw=red] (0,0) -- (0,2.3);
		\node at (0.3, 1) {$\ell_z$};
		\draw[->,draw=red] (0,0) -- (1.3,0);
		\node at (0.6,0.3) {$\ell_r$};
		\node at (-2.4,-2.8) {($\textbf{c}$)}; 
	\end{scope}
\end{tikzpicture}
\caption{Dielectric spheroidal of three different aspect ratios with $\epsilon_2$. (\textbf{a}) Dielectric sphere with radius $\ell $, (\textbf{b}) dielectric oblate spheroidal oriented along the x-axis with  semi-major axis  $\ell_r$, (\textbf{c}) dielectric oblate spheroidal with semi-major axis $\ell_z$ oriented along the z-axis, all inside a dielectric medium with $\epsilon_2$ and a point charge at the center.}\label{shapes}
\end{figure*}

In the case of a spherical dielectric cavity of radius $\ell$ and dielectric permeability $\epsilon_1$ into a dielectric medium of $\epsilon_2$, let us consider the  point charge at the center to make the calculations more comfortable as shown in Figure \ref{shapes}.\textbf{a}. The electrostatic potential in spherical coordinates  at $r$ due to a charge particle is

{\large{
		\begin{align}\label{ff}
			\phi(r)=\begin{Bmatrix}	\frac{q}{4\pi\epsilon_0}\left ( \frac{\epsilon_1-\epsilon_2}{\ell\epsilon_2\epsilon_1}+\frac{1}{r}\right )& \text{si}\quad r<\ell,\\ \frac{q}{4\pi\epsilon_0\epsilon_2r}& \text{si}\quad  r>\ell.\\ 
			\end{Bmatrix} 
		\end{align}
}}
where 
For dielectric confinement we are only interested in knowing how the external dielectric medium affects the charge on the sphere, so we only take the part of the potential referring to $r<\ell$, for the electron the matrix elements of Dielectric confinement are

\begin{align}\label{Hesf}
	\hat{H}_{e}^{Diel}=&\frac{q^2}{4\pi\epsilon_0}\left ( \frac{\epsilon_1-\epsilon_2}{\ell\epsilon_2\epsilon_1}\right)\langle \Psi_{i}(\vec{r}\,)|\Psi_l(\vec{r}\,)\rangle\notag\\=&\frac{q^2}{4\pi\epsilon_0}\left ( \frac{\epsilon_1-\epsilon_2}{\ell\epsilon_2\epsilon_1}\right)\delta_{i,l},
\end{align}

\noindent where if $\epsilon_1>\epsilon_2$
the potential will always be of the sign of the charge that produces it, thus the matrix elements are positive, for electron and hole $\hat{H}_{e}^{Diel}= \hat{H}_{h}^{Diel}.$


\subsubsection{Dielectric confinement for a prolate spheroidal}
To resolved the potential problem, we define the prolate spheroidal 

\begin{align}
&x=\frac{\alpha}{2}\sqrt{(\xi^2-1)(1-\eta^2)}\cos\varphi,\\
&y=\frac{\alpha}{2}\sqrt{(\xi^2-1)(1-\eta^2)}\sin\varphi,\\
&z=\frac{\alpha}{2}\xi\eta,
\end{align}

\noindent where $\xi\,\, \epsilon \,\, [1,\infty)$, $\eta\,\,\epsilon \,\,[-1,1]$ and $\varphi\,\,\epsilon\,\, [0,2\pi]$.\\
The potential problem inside of a prolate spheroidal has a general solution with the charge inside the sphere located at $\mathbf{r}_s=(\xi_0,\eta_0,\varphi_0)$ \cite{prolate}, which is

\begin{align}\label{Prolate}
\phi({\xi,\eta,\varphi})= & \frac{q}{4\pi\epsilon_1\epsilon_0\left|\mathbf{r}-\mathbf{r}_s\right|}+\frac{q}{4\pi\epsilon_0\epsilon_1 \alpha} \sum_{n=0}^{\infty} \sum_{m=0}^n\left(\epsilon_1-\epsilon_2\right) \notag\\&\times H_{m n} P_n^m\left(\eta_0\right) P_n^m\left(\xi_0\right) Q_n^m\left(\xi_1\right) \widehat{Q}_n^m\left(\xi_1\right)\\
&\times K_{m n}^{-1} \cos m \varphi P_n^m(\eta) P_n^m(\xi)\notag,
\end{align}

\noindent where 
\begin{align}
&K_{m n}^{-1}=2(2n+1)(2-\delta_{m0})(-1)^m \Big[\frac{(n-m)!}{(n+m)!}\Big]^2,\\& \widehat{P}_n^m\left(\xi_1\right)=(n-m+1) P_{n+1}^m\left(\xi_1\right)-(n+1) \xi_1 P_n^m\left(\xi_1\right), \\
& \widehat{Q}_n^m\left(\xi_1\right)=(n-m+1) Q_{n+1}^m\left(\xi_1\right)-(n+1) \xi_1 Q_n^m\left(\xi_1\right).
\end{align}

\noindent $Q^m_n(x)$ $P^m_n(x)$ and $Q^m_n(x)$ are the associated Legendre polynomials of the first and the second kind.\\
If the charge located at the point $\mathbf{r}_s=(0,0,0)$, we obtain azimuthal geometry thus $m=0$, so that the equation (\ref{Prolate}) is written 
\begin{align}
\phi(\xi,\eta,\varphi)=\frac{q}{4\pi\epsilon_0\epsilon_1 \alpha}\sum_{n=0}^{\infty }\frac{A_n}{B_n}P_n(\eta)P_n(\xi),
\end{align}
where 
\begin{align}
&A_n=\frac{n!(\epsilon _1-\epsilon_2)Q_n(\xi_1)(Q_{n+1}(\xi_1)-\xi_1Q_{n}(\xi_1))}{((n/2)!)^2},
\\&B_n=\epsilon_1 P_n(\xi_1)(n+1)(Q_{n+1}(\xi_1)-Q_{n}(\xi_1))\notag\\&\qquad-\epsilon_1 Q_n(\xi_1)(n+1)(P_{n+1}(\xi_1)-\xi_1P_{n}(\xi_1)),
\end{align}
thus the matrix elements are calculated as
\begin{align}\label{Hprolate}
\hat{H}_{e}^{Diel}=\frac{q^2}{4\pi\epsilon_0\epsilon_1 \alpha}\sum_{n=0}^{\infty }\frac{A_n}{B_n}\langle \Psi_{i}(\vec{r}\,)|P_n(\eta)P_n(\xi)|\Psi_l(\vec{r}\,)\rangle.
\end{align}

This equation allows us to calculate the matrix elements given a particular wave function $\Psi_{i}(\vec{r}\,)$ with $\vec{r}=(\xi,\eta,\varphi)$, for a prolate spheroidal QD, when a charged particle is located at the origin.

\subsubsection{Dielectric confinement for a oblate spheroidal}
Analogous to the prolate case we write the coordinates

\begin{align}
&x=\frac{\alpha}{2}\sqrt{(\xi^2+1)(1-\eta^2)}\cos\varphi,\\
&y=\frac{\alpha}{2}\sqrt{(\xi^2+1)(1-\eta^2)}\sin\varphi,\\
&z=\frac{\alpha}{2}\xi\eta,
\end{align}

\noindent where $\xi\,\, \epsilon \,\, [1,\infty)$, $\eta\,\,\epsilon \,\,[-1,1]$ and $\varphi\,\,\epsilon\,\, [0,2\pi]$, taking the solution of \cite{Oblato}, for the charge at $\mathbf{r}_s=(\xi_0,\eta_=,\varphi_0)$,  we obtain the electrostatic potential for the prolate system (Figure \ref{shapes}.{\textbf{c}})

\begin{align}
\phi(\xi,\eta,\varphi)=\frac{q}{4\pi\epsilon_0\epsilon_1 \alpha}\sum_{n=0}^{\infty }\frac{C_n}{D_n}P_n(\eta)P_n(i\xi),
\end{align}

\noindent where 

\begin{align}
&C_n=\frac{n!(\epsilon _1-\epsilon_2)Q_n(i\xi_1)(Q_{n+1}(i\xi_1)-i\xi_1Q_{n}(i\xi_1))}{((n/2)!)^2},
\\&D_n=\epsilon_1 P_n(i\xi_1)(n+1)(Q_{n+1}(i\xi_1)-Q_{n}(i\xi_1))\notag\\&\qquad-\epsilon_1 Q_n(\xi_1)(n+1)(P_{n+1}(i\xi_1)-i\xi_1P_{n}(i\xi_1)),
\end{align}
This equation like equation (\ref{Hprolate}) allows  us calculate the matrix elements given a particular wave function.

\noindent thus the matrix elements are calculated as

\begin{align}\label{Hoblate}
\hat{H}_{e}^{Diel}=\frac{q^2}{4\pi\epsilon_0\epsilon_1 \alpha}\sum_{n=0}^{\infty }\frac{C_n}{D_n}\langle \Psi_{i}(\vec{r}\,)|P_n(\eta)P_n(i\xi)|\Psi_l(\vec{r}\,)\rangle,
\end{align}


\begin{figure}
\centering
\includegraphics[width=0.5\textwidth]{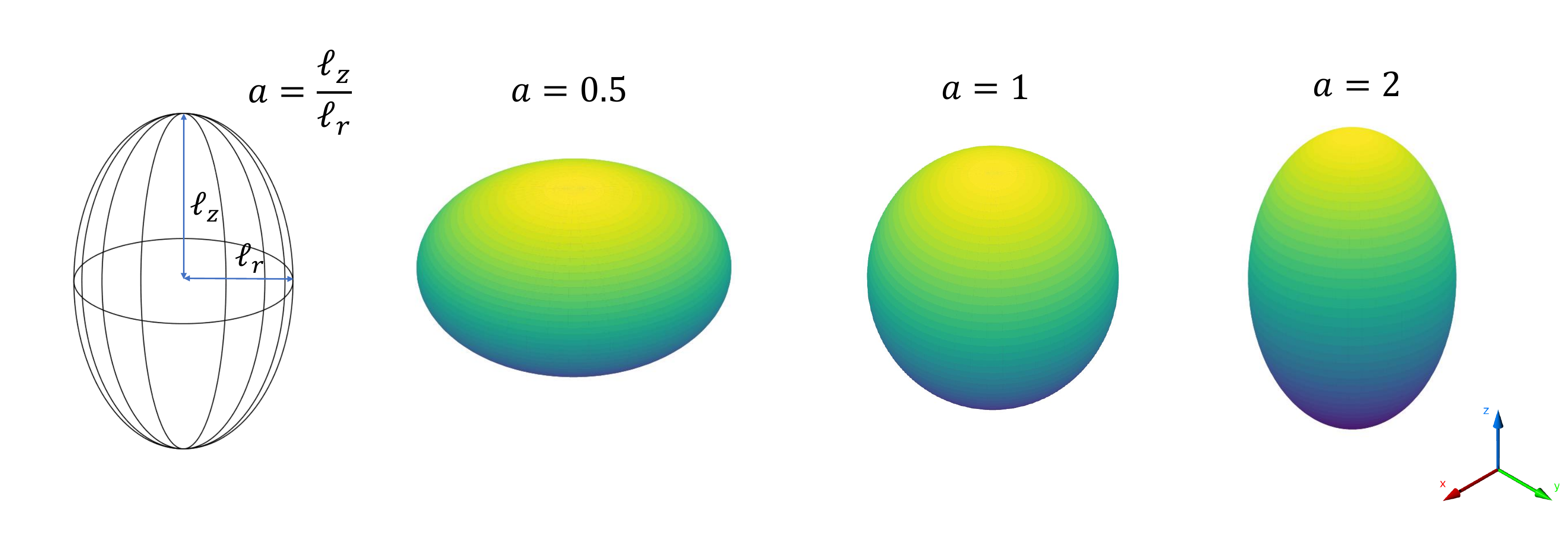}
\caption{Aspect ratio for Spheroidal quantum dots.}
\label{fig:Asp}
\end{figure}

\subsection{Kinematic and confinement part}
The Hamiltonians $\hat{H_e}$  and $\hat{H_h}$ in a three-dimensional sistem have an analogous from, which can be expressed as

\begin{align}
\hat{H}_{e/h}=\frac{\hat{P^2}}{2m_{e/h}}+\frac{1}{2}m^{e/h}\omega_x\hat{x}^2+\frac{1}{2}m^{e/h}\omega_y\hat{y}^2+\frac{1}{2}m^{e/h}\omega_z \hat{z}^2,	
\end{align}

\noindent here $\hat{P}=(\hat{P_x},\hat{P_y},\hat{P_z})$ is the momentum operator, $m^{e/h}$ is the electron-hole effective mass, $\omega_x$, $\omega_y$, and $\omega_z$ are the confinement frequencies for each direction, that in axial symmetry, $\omega_x=\omega_y=\omega_r$, so that the Hamiltonian takes the form

\begin{align}     
\hat{H}_{e/h}=\frac{\hat{P^2}}{2m^{e/h}}+\frac{1}{2}m^{e/h}\omega_r(\hat{x}^2+\hat{y}^2)+\frac{1}{2}m^{e/h}\omega_z \hat{z}^2,
\end{align}

\noindent with eigenvalues  
\begin{align}\label{Energy}	
E_{n_e,m_e,s_e,n_h,m_h,s_h}&=\hbar \omega_{r_e}(n_e+m_e+1)+\hbar \omega_{z_e}(s_e+\frac{1}{2})\notag\\&+\hbar \omega_{r_h}(n_h+m_h+1)+\hbar \omega_{z_h}(s_h+\frac{1}{2}),
\end{align}

\noindent where the quantum numbers $n_{e/h}$ and $m_{e/h}$ correspond to in-plane excitations (x-y), while $s_{e/h}$  refers to the out-of-plane (z), for the electron and hole respectively.\\
Based on the confinement frequencies, we can define the shape of the QD. In the case of $\omega_z = \omega_r$, the QD will have a spherical shape. For $\omega_z > \omega_r$ it will have an oblate form, and for $\omega_z <\omega_r$, it will exhibit a prolate geometry. This provides a very important geometrical parameter for the  QDs, known as the aspect ratio.
The characteristic in-plane and out-of-plane lengths for the electron (hole) are given by

\begin{align}\label{longitud}    
\ell_r^{e/h}=\sqrt{\frac{\hbar}{2m^{e/h}\omega_r^{e/h}}},
\end{align}

\noindent and
\begin{align}
\ell_z^{e/h}=\sqrt{\frac{\hbar}{2m^{e/h}\omega_z^{e/h}}}.
\end{align}
In this word we consider electron-hole symmetry, such that $\ell_{z/r}^{e}=\ell_{z/r}^{e}=\ell_{z/r}$, in terms of which the aspect ratio is defined as 

\begin{align}\label{Asp}
a=\frac{\ell_z}{\ell_r}=\sqrt{\frac{\omega_r}{\omega_z}},
\end{align}

\noindent in Figure (\ref{fig:Asp}) we can see three different aspects of radius for a particular volume.
\subsection{Coulomb effects}
The Hamiltonian for the interactions between electron and hole is

\begin{align}
\hat{H}_{eh}=-\sum_{ijkl}\langle\,i\,j|V|\,k\,l\rangle\hat{h}_i^\dagger\hat{e}_j^\dagger\hat{h}_k\hat{e}_l
\end{align}

\noindent where $i=\,n_{e}^{\prime}, m_{e}^{\prime}$, $s_{e}^{\prime}$, $ j=\,n_{h}^{\prime}, m_{h}^{\prime}, s_{h}^{\prime}$, $k=\,n_{h}, m_{h}, s_{h}$ and $l=\,n_{e}, m_{e}, s_{e}$, which the composite index for the quantum numbers of the confined electron and hole, $\hat{h}$ and $\hat{e}$ ($\hat{h}^\dag$ and $ \hat{e}^\dag$) are denote the creation (annihilation) operators for electron and hole. $\langle\,i\,j|V|\,k\,l\rangle$ are the coulomb matrix elements, defined as

\begin{align}
\langle i\,j\,|V(\vec{r}_{1}-\vec{r}_{2})|\,k\,l\rangle\,\equiv\, &\frac{e^2}{4 \pi \epsilon_0 \kappa } \int^{\infty}_{\infty}\int^{\infty}_{\infty} d \vec{r}_{1} d \vec{r}_{2}\psi_{i}^{*}\left(\vec{r}_{1}\right) \psi_{j}^{*}\left(\vec{r}_{2}\right)\notag \\ &\times\frac{1}{\left|\vec{r}_{1}-\vec{r}_{2}\right|} \psi_{k}\left(\vec{r}_{2}\right) \psi_{l}\left(\vec{r}_{1}\right),  
\end{align}

\noindent where $\kappa$ is the dielectric constant of the quantum dot, accounting for the screening of the Coulomb interaction within the material, and $\epsilon_0$ is the vacuum permittivity. Explicitly, the Coulomb matrix elements are,
\newpage
{\begin{strip}
	\begin{align}\label{A1}
		\langle \,n_{e}^{\prime} m_{e}^{\prime} s_{e}^{\prime} n_{h}^{\prime} m_{h}^{\prime}   s_{h}^{\prime}|V|n_{h} m_{h} s_{h}n_{e}m_{e} s_{e}&\rangle=\frac{e^2}{8 \pi^2 \epsilon_0 \kappa }\frac{\delta_{m\prime_e+n_e+m\prime_h+n_h,\,m_h+m_e+n\prime_h+n\prime_e}\delta_{s_e+s_h+s\prime_e+s\prime_h,\,even} }{\ell_r\sqrt{n\prime_e!\,m\prime_e!\,s\prime_e!\,n_e!\,m_e!\,s_e!\,n\prime_h!\,m\prime_h!\,s\prime_h!\,n_h!\,m_h!\,s_h!}}\notag\\&\times\sum_{P_1=0}^{min\left({n\prime_e,n_e}\right)}\sum_{P_2=0}^{min\left({m\prime_e,m_e}\right)}\sum_{P_3=0}^{min\left({s\prime_e,s_e}\right)}\sum_{P_4=0}^{min\left({n\prime_h,n_h}\right)}\sum_{P_5=0}^{min\left({m\prime_h,m_h}\right)}\sum_{P_6=0}^{min\left({s\prime_h,s_h}\right)} P_1!\,P_2!\,P_3!\,P_4!\,P_5!\,P_6!\notag\\ &\times\displaystyle\binom{n\prime_e}{P_e}\displaystyle\binom{n_e}{P_e}\displaystyle\binom{m\prime_e}{P_2}\displaystyle\binom{m_e}{P_2}\displaystyle\binom{s\prime_e}{P_3}\displaystyle\binom{s_e}{P_3}\displaystyle\binom{n\prime_h}{P_4}\displaystyle\binom{n_h}{P_4}\displaystyle\binom{m\prime_h}{P_5}\displaystyle\binom{m_h}{P_5}\displaystyle\binom{s\prime_h}{P_6}\displaystyle\binom{s_h}{P_6}\notag\\ & \times \frac{1}{2^{u}}\Bigr( \frac{1}{a^2}\Bigl) ^{u+\frac{1}{2}}(-1)^{\frac{1}{2}v+u+n\prime_h+m\prime_h+s\prime_h+n_h+m_h+s_h}\frac{\Gamma\left(\frac{1+2 u+v}{2}\right) \Gamma(1+u) \Gamma\left(\frac{1+v}{2}\right)}{\Gamma\left(\frac{3+2 u+v}{2}\right)}\notag\\ &\times{}_{2}F_{1}\left(u+1, \frac{1+2 u+v}{2} ; \frac{3+2 u+v}{2} ; 1-\frac{1}{a^2}\right),
	\end{align}
	\end{strip}}
	
	\noindent where the Dirac deltas $\delta_{m\prime_e+n_e+m\prime_h+n_h,\,m_h+m_e+n\prime_h+n\prime_e}$ and $\delta_{s_e+s_h+s\prime_e+s\prime_h,\,even}$ ensure conservation of the $z$-component of angular momentum and parity, respectively. $u=m\prime_e+m\prime_h+n_e+n_h-(P_1+P_2+P_4+P_5)$ and $v=s\prime_e+s_e+s\prime_h+s_h-2(P_3+P_6)$. $\Gamma(x)$ and $_2F_1(x1, x2, x3, x4)$ represent the Euler Gamma and Hypergeometric $_2F_1$ functions, respectively \cite{chen}. This expression is valid for $a \geq 1$. For $a < 1$, we need to replace the expression,
	
\begin{align}
{}_{2}F_{1}\left(u+1, \frac{1+2 u+v}{2} ; \frac{3+2 u+v}{2} ; 1-\frac{1}{a^2}\right)
\end{align}

for

\begin{align}
\Bigr(\frac{1}{a^2}\Bigl)^{-1+2u+v/2}{}_{2}F_{1}\left(\frac{1+v}{2}, \frac{1+2 u+v}{2} ; \frac{3+2 u+v}{2} ; 1-a^2\right).
\end{align}


\section{analysis}
\label{sec:analysis}
To show all effects on the QDs, we perform calculations for the ground energy and the first excited states of the  CdSe, CdTe and ZnTe quantum dots with confinement radios of 3, 5, and 7 nm (corresponding to the spherical case), and aspect ratios between 0.5 (oblate) and 2 (prolate). In these calculations, we use the material bulk parameters  shown in Table (\ref{tabla}) and we take $\epsilon_1=\epsilon_0$

Its important see that the confinement volume of the spheroidal potential is given by
\begin{align}\label{vol}
	V_c=\frac{4 \pi}{3}\ell_z \ell_r^2,
\end{align}
and the superficial area is,
\begin{align}
	A_s &= 2\pi \ell_r^2 (1+\frac{1-\varrho^2}{\varrho}\tanh^{-1}[\varrho])\\ \text{with}&\qquad \varrho^2=1-a^2\quad\text{if}\quad a<1,\notag\\
	A_s &= 2\pi \ell_r^2 (1+\frac{a}{\varrho}\sin^{-1}[\varrho])\\ \text{with}&\qquad \varrho^2=1-a^{-2}\quad\text{if}\quad a>1,\notag
\end{align}
where $\varrho$ is the eccentricity of the ellipse formed by the cross section along the x-z plane.

\begin{table}
	\begin{center}
		\begin{tabular}{| l | c | c | c |}
			\hline
			& CdSe & CdTe & ZnTe\\ \hline
			Dielectric Constant & 10.6 & 10.4 & 9.7\\
			Energy gap (eV) & 1.75 & 1.61 & 2.2 \\
			Hole effective mass & 0.4 $m_0$ &  0.4 $m_0$ & 0.2 $m_0$  \\
			Electron effective mass & 0.13 $m_0$ & 0.096 $m_0$ & 0.15 $m_0$ \\ \hline
		\end{tabular}
		\caption{Material parameter (\(m_0\) is the electron mass){\citet{Cdse1,Cdse2}}.}
		\label{tabla}
	\end{center}
\end{table}

\subsubsection{Hamiltonian}

To perform the computational calculations, we need to construct the Hamiltonian given by the equation (\ref{H}) of size $N\times N$  where $N$ represents the number of combinations given a specific number of excitations in the system. It is essential to introduce several new variables that define the size of the matrix. These variables are $ L_{n_e}^{max}, L_{m_e}^{max}, L_{s_e}^{max}, L_{n_h}^{max}, L_{m_h}^{max}$ and $ L_{s_h}^{max}$, representing the maximum number of excitations for each corresponding quantum number.

For the each quantum number we have a number's excitations, if the excitations are between the ground and the first excited state (0, 1) , then $ L_{(n,m,s)_{e/h}}^{max} $ is 2, for  excitations between the ground and the second excited state (0, 1, 2), $ L_{(n,m,s)_{e/h}}^{max} $ is 3, and  $ L_{(n,m,s)_{e/h}}^{max} $ is 4 when the excitation are between the ground state and the third excited state (0, 1, 2, 3),
Thus, the total number of combinations, \( N \), is determined by the product of these \( L \) values, reflecting the total possible states for the quantum system given the specified excitation levels as
\begin{align*}
	N=Ln_{e}^{max}Lm_{e}^{max}Ls_{e}^{max}Ln_{h}^{max}Lm_{h}^{max}Ls_{h}^{max},
\end{align*}
For each quantum number, the excitation ranges are labeled as
\begin{itemize}
	\item For 0 (\textbf{s}), \(  L_{(n,m,s)_{e/h}}^{max}  = 1 \) resulting in \( N_s = 1 \).
	\item For 0-1 (\textbf{s,p)}, \(L_{(n,m,s)_{e/h}}^{max} = 2 \) resulting in \( N_{sp} = 2^6 = 64 \).
	\item For 0-2 (\textbf{s,p,d)}, \( L_{(n,m,s)_{e/h}}^{max} = 3 \) resulting in \( N_{spd} = 3^6 = 729 \).
	\item For 0-3 (\textbf{s,p,d,f)}, \(L_{(n,m,s)_{e/h}}^{max}= 4 \) resulting in \( N_{spdf} = 4^6 = 4096 \).
\end{itemize}
\begin{figure}
	
	\begin{tikzpicture}
		\node[anchor=south west,inner sep=0] (image) at (0,0){\includegraphics[width=0.5\textwidth]{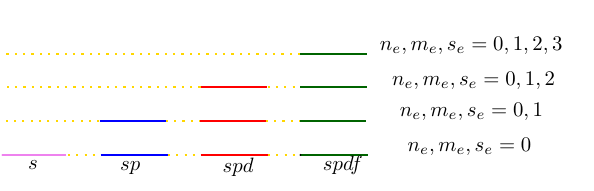}};
	\end{tikzpicture}
	\caption{$ N=Ln_{e}^{max}Lm_{e}^{max}Ls_{e}^{max}Ln_{h}^{max}Lm_{h}^{max}Ls_{h}^{max}$}
	\label{Spsf}
\end{figure}
The Figure (\ref{Spsf}) illustrates the labels and their corresponding values for each energy-stated.
\subsubsection{Energy Ground-first Stated shape}
To compare our computational results we can write the ground state energy of the system without considering Coulomb effects to see the aspect ratio form, knowing that for such state, we have $n_e=m_e=s_e=n_h=m_h=s_h=0$, so that the equation (\ref{Energy}) takes the form
\begin{align}
	E_{0,0,0,0,0,0}=\hbar \omega_{r_e}+\frac{1}{2}\hbar \omega_{z_e}+\hbar \omega_{r_h} +\frac{1}{2}\hbar \omega_{z_h}.
\end{align}
The confining frequencies can be written in terms of the corresponding lengths as $\omega_{r/z}^{e/h}=\frac{\hbar}{2m_{e/h}\ell_{r/z}}^2$. Thus
\begin{align}
	E_{0,0,0,0,0,0}=&\frac{\hbar^2}{2m^{e}\ell_{r}^2}+\frac{1}{2}\frac{\hbar^2}{2m^{e}\ell_{z}^2}+ \frac{\hbar^2}{2m^{h}\ell_{z}^2}+\frac{1}{2}\frac{\hbar^2}{2m^{h}\ell_{z}^2}\notag\\E_{0,0,0,0,0,0}=&\frac{\hbar^2}{2m^{e}}\Big(\frac{1}{\ell_r^2}+\frac{1}{2\ell_{z}^2}\Big)+ \frac{\hbar^2}{2m^{h}}\Big(\frac{1}{\ell_r^2}+\frac{1}{2\ell_{z}^2}\Big)\\ E_{0,0,0,0,0,0}=&\frac{\hbar^2}{2}\Big(\frac{1}{m^e}+\frac{1}{m_h}\Big)\Big(\frac{1}{\ell_r^2}+\frac{1}{2\ell_{z}^2}\Big)\notag.
\end{align}
From equations (\ref{Asp}) and (\ref{vol}), we can write the characteristic lengths in terms of the aspect ratio, at given confined volume $V_c$. Hence, we have
\begin{align}
	\ell_r^2 = \Bigr(\frac{3 V_c}{4 \pi a}\Bigl)^{2/3}\quad \text{and}\quad \ell_z^2 = \Bigr( \frac{3 V_ca^2}{4 \pi }\Bigl)^{2/3},
\end{align}

thus

\begin{align}\label{E000}
	E_{0,0,0,0,0,0}=&\frac{\hbar^2}{2}\Big(\frac{1}{m^e}+\frac{1}{m_h}\Big)\Big(\Big(\frac{4\pi}{3Vc}\Big)^{2/3}\sqrt[3]{a^2} +\Big(\frac{4\pi}{3Vc}\Big)^{2/3}\frac{1}{2\sqrt[3]{a^4}}\Big)
	\notag\\ E_{0,0,0,0,0,0}=&\frac{\hbar^2}{2}\Big(\frac{4\pi}{3Vc}\Big)^{2/3} \Big(\frac{1}{m^e}+\frac{1}{m_h}\Big)\Big(a^{2/3}+\frac{1}{2} a^{-4/3}\Big),
\end{align}

\begin{figure*}[H]
	\begin{tikzpicture}
		\node[anchor=south west] at (0, 0) {\includegraphics[width=0.9\textwidth]{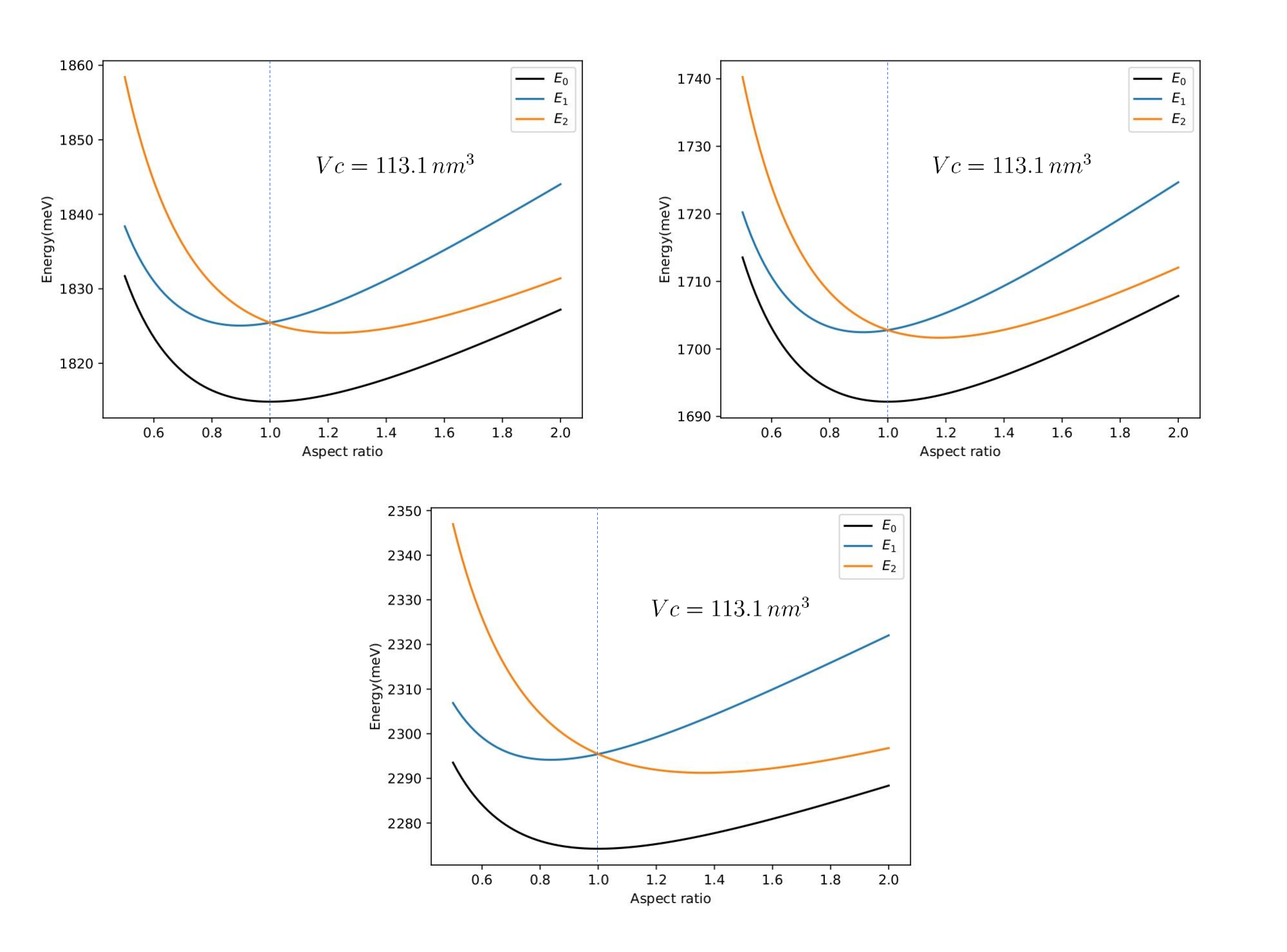}};
		\node at (4.3, 11.5) {CdSe}; 
		\node at (12, 11.5) {CdTe};
		\node at (8.3,6) {ZnTe};  
		\node at (0.8,6.7) {$a)$}; 
		\node at (8.3,6.7) {$b)$}; 
		\node at (4.5,1) {$c)$}; 
	\end{tikzpicture}	
	\caption{Analytical ground, first excited and second excited states for a) CdSe b) CdTe and c) ZnTe with $Vc=113.1\,nm^3$.}
	\label{fig:enter-3}
\end{figure*}

As a function of the aspect ratio, equation (\ref{E000}) has a minimum at $a=1$. This provides a verifiable benchmark for the numerical results obtained from the written computational code. Analogously we can calculate for the first and second excited state, which will be characterized by $n_e=m_e=s_e=m_h=s_h=0$,$n_h=1$ ($n_e=m_e=s_e=n_h=s_h=0$, $m_h=1$ since it is degenerate in the plane),  and $n_e=m_e=s_e=m_h=n_h=0$, $s_h=1$ , for the oblate and prolate cases, respectively. Figure \ref{fig:enter-3} shows these theoretical curves for a confining volume $Vc=113.1\,nm^3$ in the 3 studied  materials.

\subsubsection{Dielectrical confinement}
In the section (\ref{sec:Dielectric}) we made a first approximation to quantify the effects of dielectric confinement or self-energy, we obtain the expressions ((\ref{Hesf}, \ref{Hprolate}), (\ref{Hoblate})), however we must clarify that this approximation has several aspects to improve, such as the location of both charges at the origin, which would imply that the charges have an infinite coulomb potential, which we know is erroneous.

If we evaluate equation (\ref{Hesf}) for CdSe in the vacuum for a electron, we obtain that self-energy is $433.6\,\, meV$, which is a significant value close to the half of the energy ground state in computational calculations.
\subsubsection{Computational results}
Figures (\ref{fig:three_images}) show the results for the ground state within each of the used levels of approximation and two expected features can be observed:
\begin{itemize}
	\item The smaller the confined volume, the higher the obtained eigenenergies.
	\item The energy difference between levels of approximation is reduced as the used basis is enlarged. Thus, the exact energy values should be very close to the curves computed with s,p,d and f levels. We refer to it as the full 			configuration interaction (FCI) result and will focus on that level of 				approximation for the rest of the simulations.   
\end{itemize}
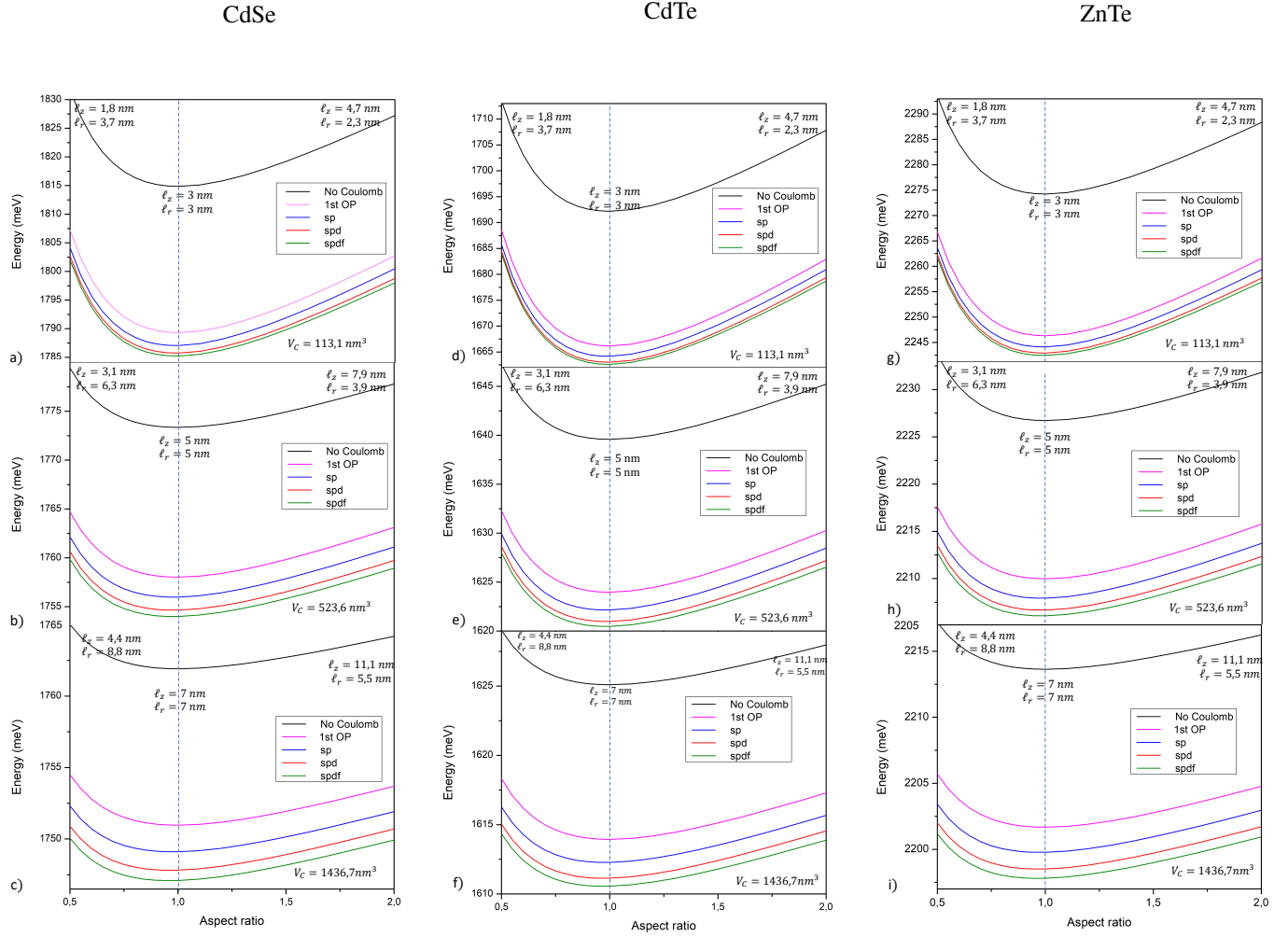
\begin{figure*}
	\centering
	\begin{tikzpicture}
		\node[anchor=south west] at (-1.3, 0) {\includegraphics[width=0.38\textwidth]{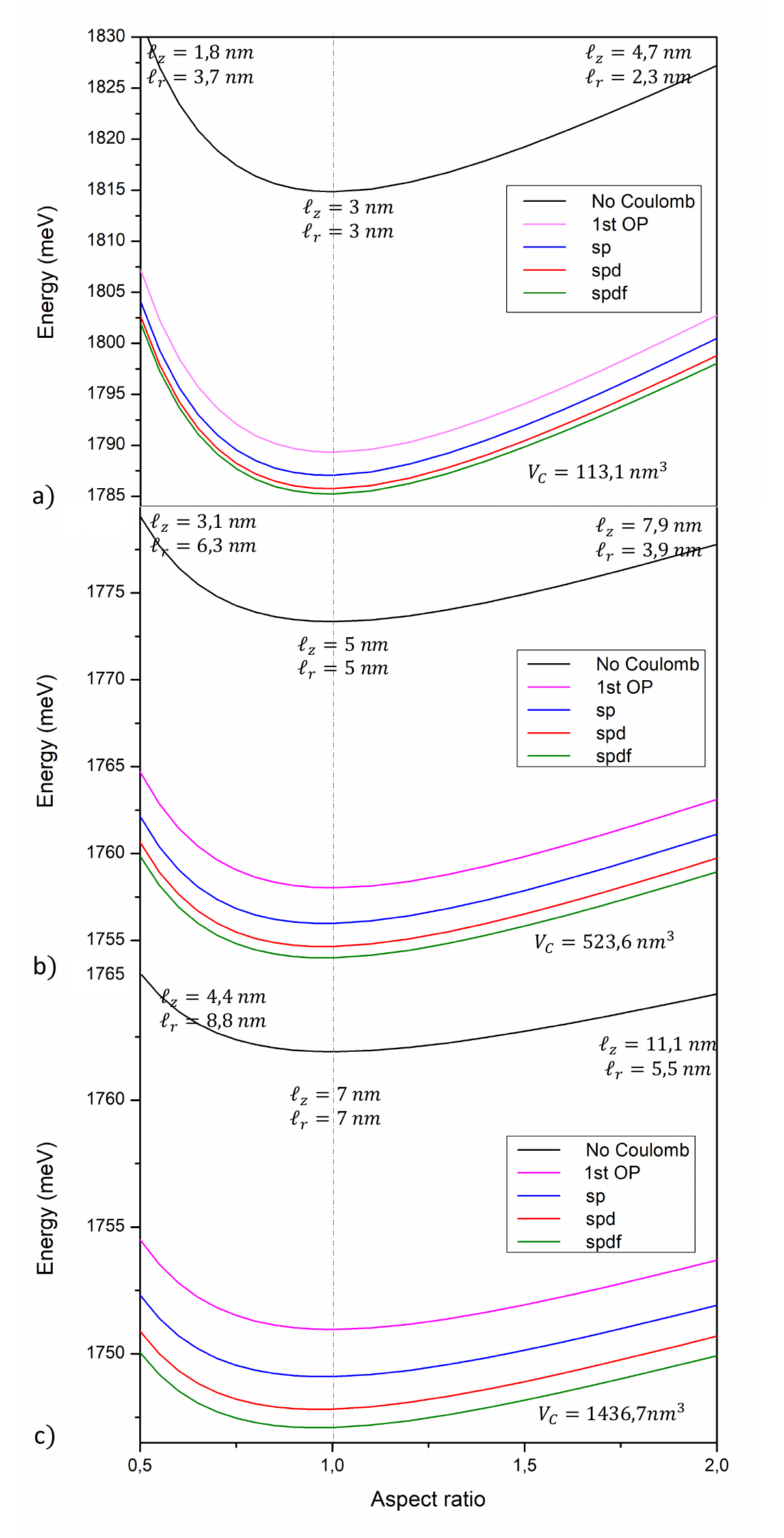}};
		\node at (2.5, 13.35) {CdSe}; 
		\node[anchor=south west] at (5, 0) {\includegraphics[width=0.38\textwidth]{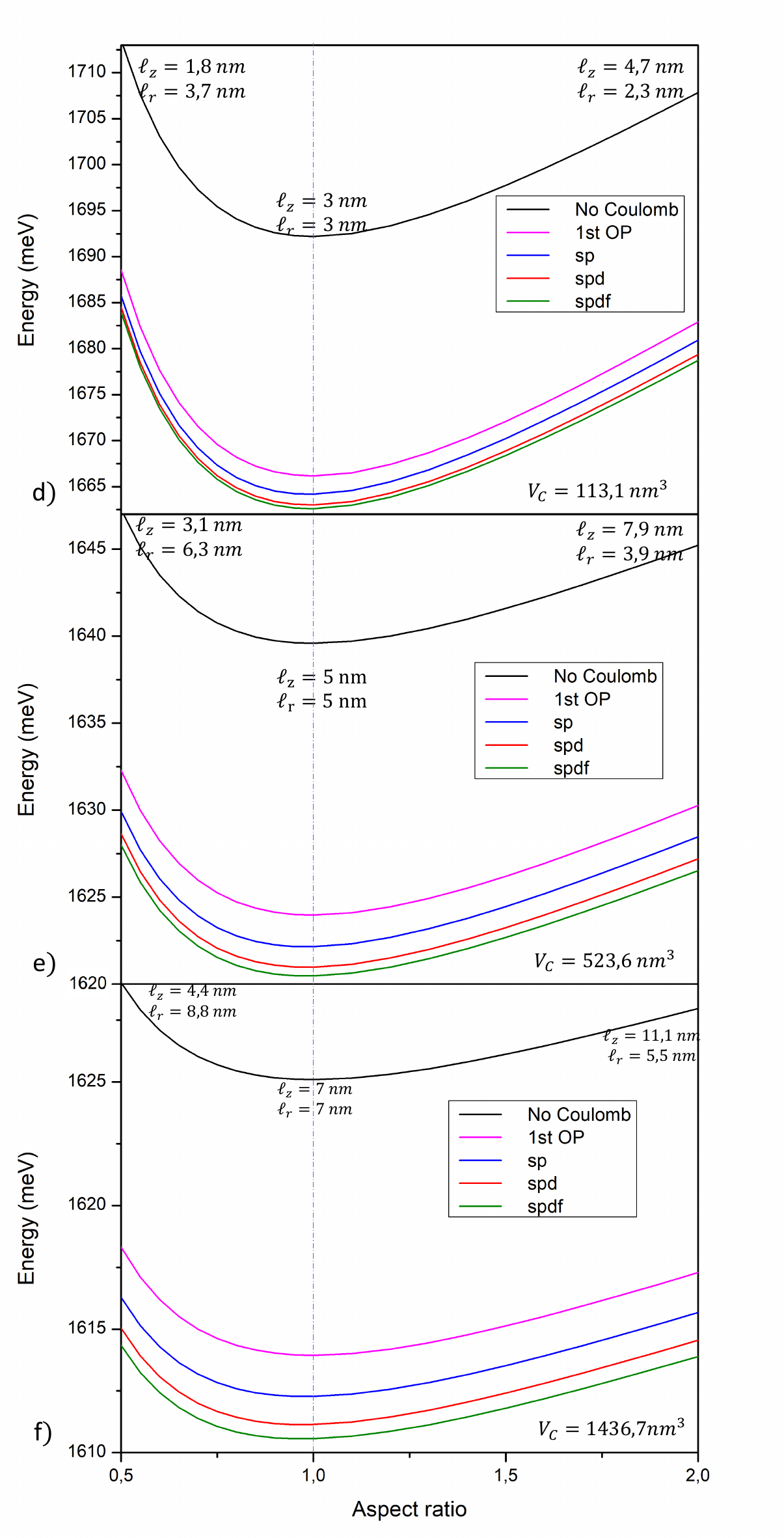}};
		\node at (8.5, 13.4) {CdTe}; 
		\node[anchor=south west] at (11.2, 0) {\includegraphics[width=0.38\textwidth]{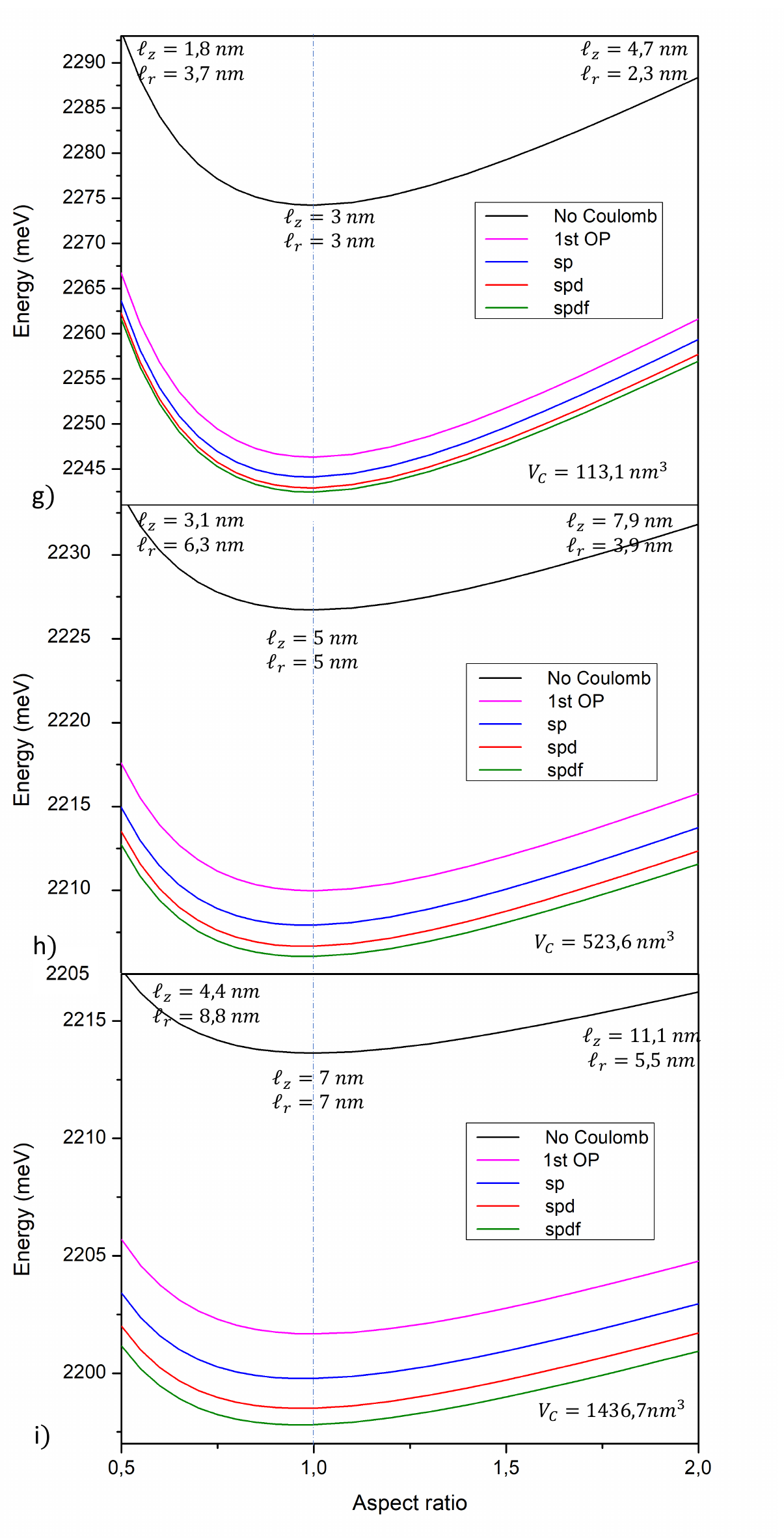}};
		\node at (14.7, 13.35) {ZnTe}; 
	\end{tikzpicture}
	\caption{ Coulomb effects on the ground-state energy for a CdSe,CdTe and ZnTe quantum dots with a confinement radius of {\textbf{ a), d), g)}} $3nm$,{\textbf{b), e), h)}} $5nm$ and {\textbf{c), f), i)}} $7nm$. Including the values of $\ell_r$ and $\ell_z$ at $a = 0.5$, $a = 1$ and $a = 2$.}
	\label{fig:three_images}
\end{figure*}
\begin{figure*}
	\centering
	\begin{tikzpicture}
		\node[anchor=south west] at (-1.3, 0) {\includegraphics[width=0.38\textwidth]{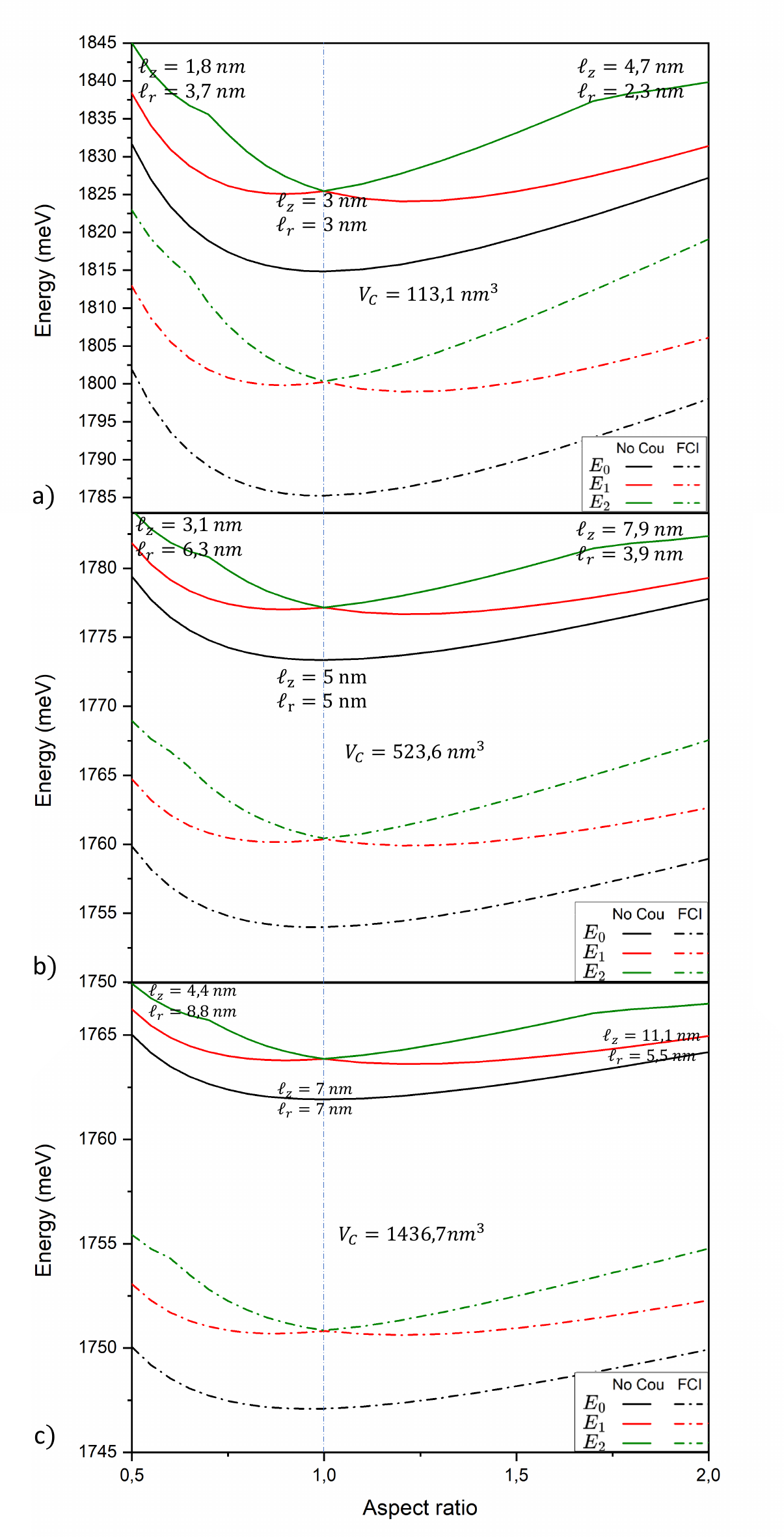}};
		\node at (2.5, 13.35) {CdSe}; 
		\node[anchor=south west] at (5, 0) {\includegraphics[width=0.38\textwidth]{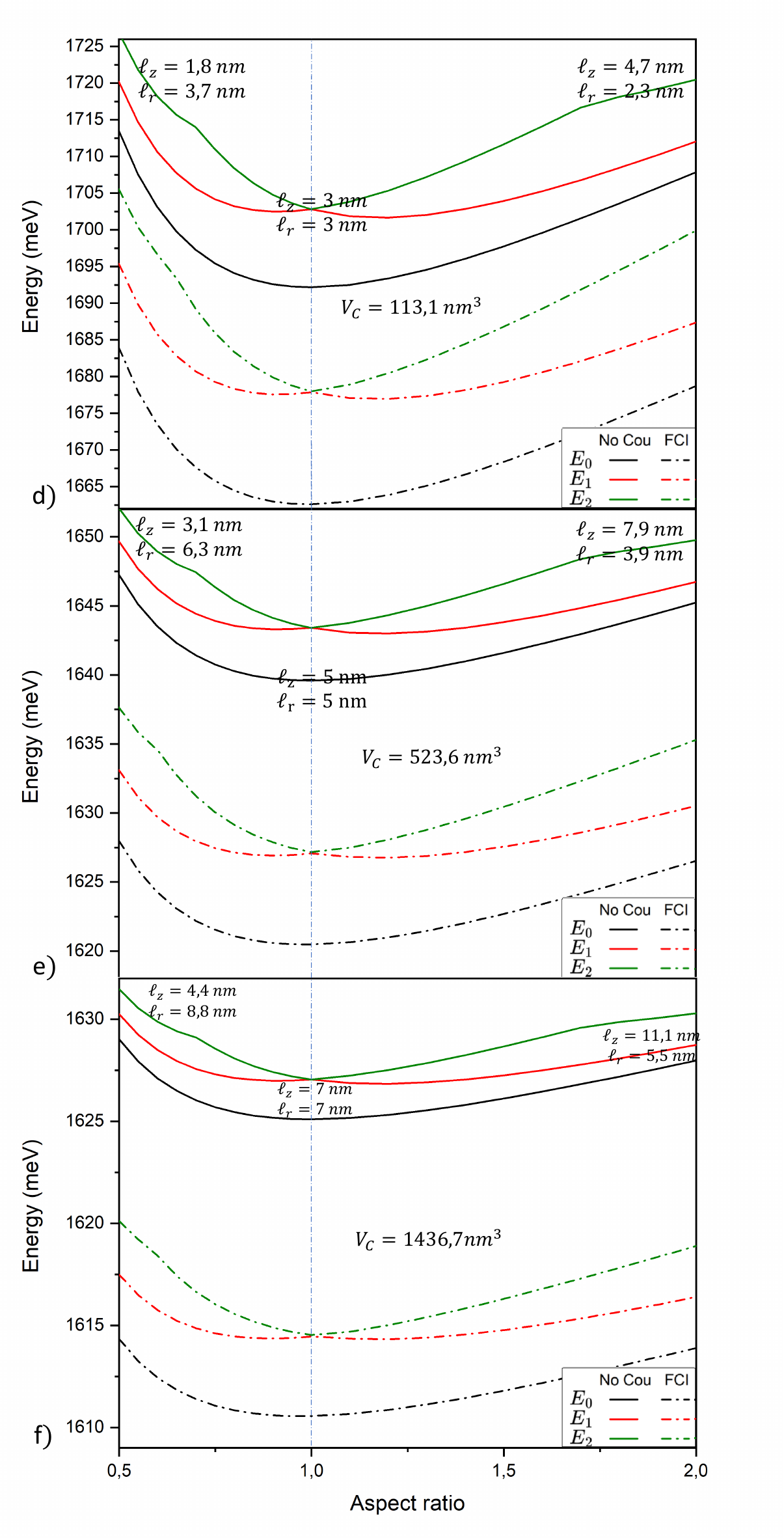}};
		\node at (8.5, 13.4) {CdTe}; 
		\node[anchor=south west] at (11.2, 0) {\includegraphics[width=0.38\textwidth]{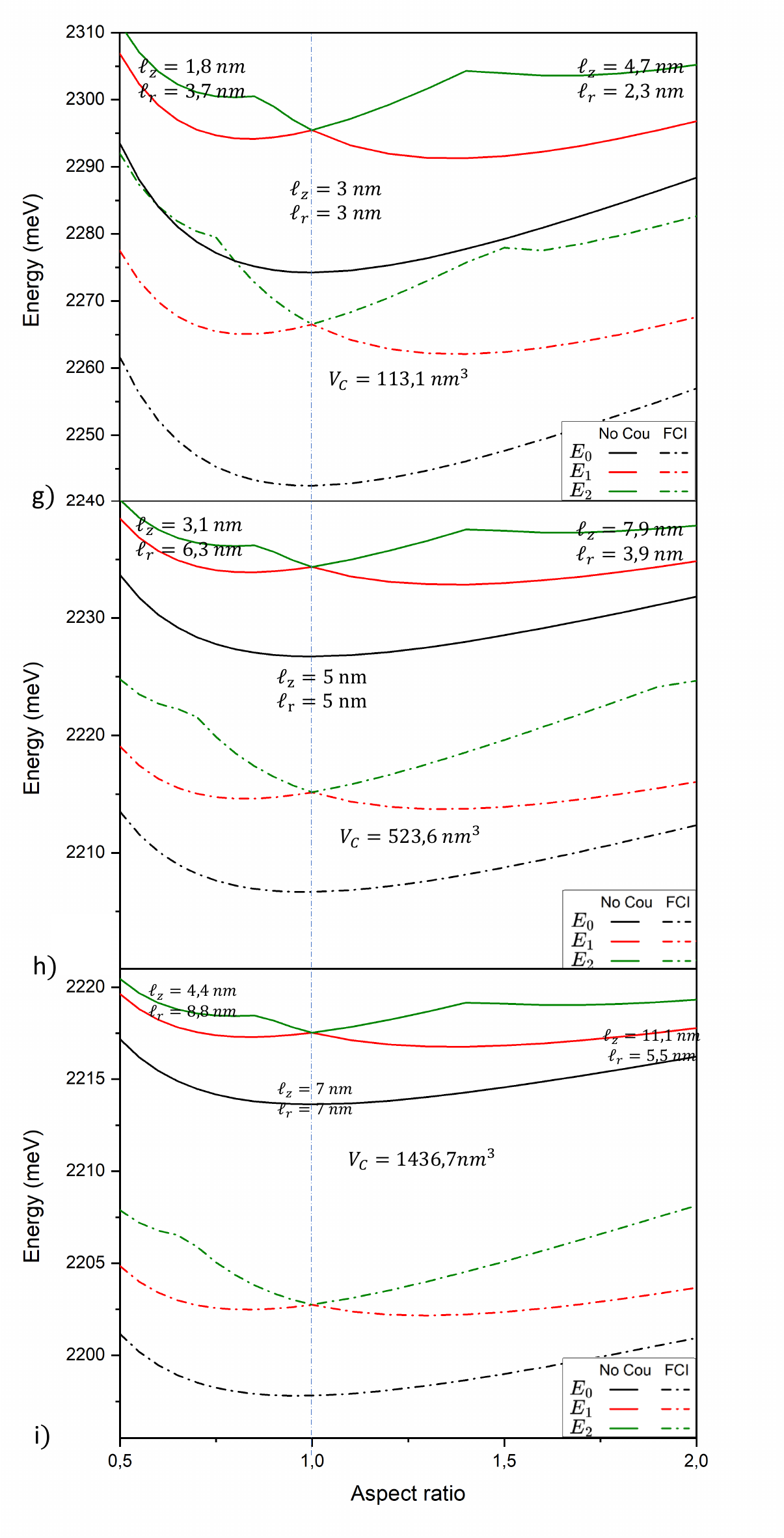}};
		\node at (14.7, 13.35) {ZnTe}; 
	\end{tikzpicture}
	\caption{ Non interacting and interacting ground, first excited and second excited states of a trapped electron hole pair in three confinement of {\textbf{ a),d),g)}} $3nm$,{\textbf{b),e),h)}} $5nm$ and {\textbf{c),f),i)}} $7nm$. Including the values of $\ell_r$ and $\ell_z$ at $a = 0.5$, $a = 1$ and $a = 2$.}
	\label{fig:Excited}
\end{figure*}
To analyze the different levels of approximation, focusing on the ground state ($E_0$), we define the following parameters
\begin{align}\label{CCC1}
	E_{0c}^{FCI}(\%E_0)=\frac{E_0-E_{0}^{spdf}}{E_0}\times 100,
\end{align}
\begin{align}\label{CCC2}
	E_{0c}^{sp}(\%E_0^{FCI})=\frac{E_{0}^{sp}}{E_0^{FCI}}\times 100,
\end{align}
\begin{align}\label{CCC3}
	E_{0c}^{spd}(\%E_0^{FCI})=\frac{E_0^{spd}}{E_0^{FCI}}\times 100,
\end{align}
\begin{align}\label{CCC4}
	E_{0c}^{1OP}(\%E_0^{FCI})=\frac{E_0^{1OP}}{E_0^{FCI}}\times 100.
\end{align}
In order to evaluate the coulomb effects in the ground state for each material and the three confining volumes on which we carry out calculations, we have compiled the quantities defined in equations (\ref{CCC1}-\ref{CCC4}) for different radius aspects in the following tables.

\begin{table}
	\resizebox{0.5\textwidth}{!} {
		\begin{tabular}{|c|c|c|c|c|c|}
			\hline
			Aspect Ratio & $E_0$(meV) & $E_{0c}^{FCI}(\%E_0) $& $E_{0c}^{sp}(\%E_{0c}^{FCI})$ & $E_{0c}^{spd}(\%E_{0c}^{FCI})$ & $E_{0c}^{1OP}(\%E_{0c}^{FCI})$ \\ \hline
			0.5 & 1831.71 & 1.62 & 92.78 & 97.73 & 82.38  \\ \hline
			1 & 1814.85 & 1.63 & 93.83 & 98.29 & 86.23  \\ \hline
			2 & 1827.21 & 1.60 & 91.60 & 97.33 & 83.82  \\ \hline
	\end{tabular}}
	\caption{Ground-state energy of CdSe for different approximations and confined volume $V_c=113.1\,\,nm^3$.}
	\label{Tabla2}
\end{table}
\begin{table}
	\centering
	\resizebox{0.5\textwidth}{!} {
		\begin{tabular}{|c|c|c|c|c|c|}
			\hline
			Aspect Ratio & $E_0$(meV) & $E_{0c}^{FCI}(\%E_0) $& $E_{0c}^{sp}(\%E_{0c}^{FCI})$ & $E_{0c}^{spd}(\%E_{0c}^{FCI})$ & $E_{0c}^{1OP}(\%E_{0c}^{FCI})$ \\ \hline
			0.5 & 1779.42 & 1.10 & 88.28 & 95.89 & 75.17  \\ \hline
			1 & 1773.35 & 1.09 & 89.78 & 96.66 & 79.21  \\ \hline
			2 & 1777.80 & 1.06 & 88.47 & 95.82 & 77.86  \\ \hline  
	\end{tabular}}
	\caption{Ground-state energy of CdSe for different approximations and confined volume $V_c=523.6\,\,nm^3$.}
\end{table}
\begin{table}
	\centering
	\resizebox{0.5\textwidth}{!} {
		\begin{tabular}{|c|c|c|c|c|c|}
			\hline
			Aspect Ratio & $E_0$(meV) & $E_{0c}^{FCI}(\%E_0) $& $E_{0c}^{sp}(\%E_{0c}^{FCI})$ & $E_{0c}^{spd}(\%E_{0c}^{FCI})$ & $E_{0c}^{1OP}(\%E_{0c}^{FCI})$ \\ \hline
			0.5 & 1765.01 & 0.85 & 84.86 & 94.34 & 70.23  \\ \hline
			1 & 1761.91 & 0.84 & 86.43 & 95.11 & 73.89  \\ \hline
			2 & 1764.18 & 0.81 & 86.03 & 94.59 & 73.56  \\ \hline
	\end{tabular}}
	\caption{Ground-state energy of CdSe for different approximations and confined volume $V_c=1436.7\,\,nm^3$.}
\end{table}
\begin{table}
	\centering
	\resizebox{0.5\textwidth}{!} {
		\begin{tabular}{|c|c|c|c|c|c|}
			\hline
			Aspect Ratio & $E_0$(meV) & $E_{0c}^{FCI}(\%E_0) $& $E_{0c}^{sp}(\%E_{0c}^{FCI})$ & $E_{0c}^{spd}(\%E_{0c}^{FCI})$ & $E_{0c}^{1OP}(\%E_{0c}^{FCI})$ \\ \hline
			0.5 & 1713.55 & 1.73 & 93.93 & 98.03 & 84.05  \\ \hline
			1 & 1692.19 & 1.75 & 94.72 & 98.53 & 87.82  \\ \hline
			2 & 1707.85 & 1.71 & 92.41 & 97.58 & 85.09  \\ \hline
	\end{tabular}}
	\caption{Ground-state energy of CdTe for different approximations and confined volume $V_c=113.1\,\,nm^3$.}
\end{table}
\begin{table}
	\centering
	\resizebox{0.5\textwidth}{!} {
		\begin{tabular}{|c|c|c|c|c|c|}
			\hline
			Aspect Ratio & $E_0$(meV) & $E_{0c}^{FCI}(\%E_0) $& $E_{0c}^{sp}(\%E_{0c}^{FCI})$ & $E_{0c}^{spd}(\%E_{0c}^{FCI})$ & $E_{0c}^{1OP}(\%E_{0c}^{FCI})$ \\ \hline
			0.5 & 1647.28 & 1.17 & 89.94 & 96.63 & 77.74  \\ \hline
			1 & 1639.59 & 1.16 & 91.24 & 97.35 & 81.72  \\ \hline
			2 & 1645.23 & 1.14 & 89.60 & 96.42 & 80.01  \\ \hline
	\end{tabular}}
	\caption{Ground-state energy of CdTe for different approximations and confined volume $V_c=523.6\,\,nm^3$.}
\end{table}
\begin{table}
	\centering
	\resizebox{0.5\textwidth}{!} {
		\begin{tabular}{|c|c|c|c|c|c|}
			\hline
			Aspect Ratio & $E_0$(meV) & $E_{0c}^{FCI}(\%E_0) $& $E_{0c}^{sp}(\%E_{0c}^{FCI})$ & $E_{0c}^{spd}(\%E_{0c}^{FCI})$ & $E_{0c}^{1OP}(\%E_{0c}^{FCI})$ \\ \hline 0.5 & 1629.02 & 0.90 & 86.71 & 95.24 & 72.92  \\ \hline 1 & 1625.10 & 0.89 & 88.22 & 96.02 & 76.76  \\ \hline 2 & 1627.97 & 0.87 & 87.36 & 95.30 & 75.88  \\ \hline
	\end{tabular}}
	\caption{Ground-state energy of CdTe for different approximations and confined volume $V_c=1436.7\,\,nm^3$.}
\end{table}
\begin{table}
	\centering
	\resizebox{0.5\textwidth}{!} {
		\begin{tabular}{|c|c|c|c|c|c|}
			\hline
			Aspect Ratio & $E_0$(meV) & $E_{0c}^{FCI}(\%E_0) $& $E_{0c}^{sp}(\%E_{0c}^{FCI})$ & $E_{0c}^{spd}(\%E_{0c}^{FCI})$ & $E_{0c}^{1OP}(\%E_{0c}^{FCI})$ \\ \hline
			0.5 & 2293.53 & 1.39 & 93.71 & 98.03 & 84.05  \\ \hline
			1 & 2274.23 & 1.40 & 94.75 & 98.53 & 87.82  \\ \hline
			2 & 2288.38 & 1.37 & 92.36 & 97.58 & 85.09  \\ \hline
	\end{tabular}}
	\caption{Ground-state energy of ZnTe for different approximations and confined volume $V_c=113.1\,\,nm^3$.}
\end{table}
\begin{table}
	\centering
	\resizebox{0.5\textwidth}{!} {
		\begin{tabular}{|c|c|c|c|c|c|}
			\hline
			Aspect Ratio & $E_0$(meV) & $E_{0c}^{FCI}(\%E_0) $& $E_{0c}^{sp}(\%E_{0c}^{FCI})$ & $E_{0c}^{spd}(\%E_{0c}^{FCI})$ & $E_{0c}^{1OP}(\%E_{0c}^{FCI})$ \\ \hline  0.5 & 2233.67 & 0.90 & 92.83 & 96.31 & 79.76  \\ \hline
			1 & 2226.72 & 0.90 & 93.71 & 97.06 & 83.48  \\ \hline
			2 & 2231.82 & 0.87 & 92.86 & 96.14 & 82.41  \\ \hline
	\end{tabular}}
	\caption{Ground-state energy of ZnTe for different approximations and confined volume $V_c=523.6\,\,nm^3$.}
\end{table}
\begin{table}
	\centering
	\resizebox{0.5\textwidth}{!} {
		\begin{tabular}{|c|c|c|c|c|c|}
			\hline
			Aspect Ratio & $E_0$(meV) & $E_{0c}^{FCI}(\%E_0) $& $E_{0c}^{sp}(\%E_{0c}^{FCI})$ & $E_{0c}^{spd}(\%E_{0c}^{FCI})$ & $E_{0c}^{1OP}(\%E_{0c}^{FCI})$ \\ \hline         0.5 & 2217.18 & 0.72 & 85.94 & 94.77 & 71.68  \\ \hline 1 & 2213.63 & 0.71 & 87.51 & 95.54 & 75.30  \\ \hline 2 & 2216.23 & 0.69 & 86.84 & 94.94 & 74.92  \\ \hline
	\end{tabular}}
	\caption{Ground-state energy of ZnTe for different approximations and confined volume $V_c=1436.7\,\,nm^3$.}
	\label{Tabla10}
\end{table}
In Tables (\ref{Tabla2}) to (\ref{Tabla10}), we see:
\begin{itemize}
	\item For the three materials with $V_c=113.1\,\,nm^3$, there is a greater contribution of Coulomb effects in the case of an aspect ratio of 1 (spherical). for $V_c=523.6\,\,nm^3$, and $V_c=1436.7\,\,nm^3$, the greatest contribution is found in the case of $a=0.5$.
	\item As the volume increases, the Coulomb contribution in the ground state decreases, and the percentage contribution in each of the approximations becomes smaller.
	\item The approximation levels are better for an aspect ratio of 1, regardless of the material and volume.
	\item The material in which the Coulomb contribution is greater is CdTe, while the material with the lowest contribution is ZnTe.
	\item One could argue that the first-order perturbation results provide a good approximation for smaller confinement volumes, as they are above $80\%$.
\end{itemize}
From the above results, we can see that the Coulomb correlations are more significant for smaller confining volumes. Therefore, we will now analyze the ground, first and second excited states for confining volumes of $V_c=113.1\,\,nm^3$ in order to simplify the tables to be analyzed. The parameters we will now analyze are
\begin{align}\label{CCC5}
	E_{1c}^{FCI}(\%E_1)=\frac{E_1-E_{1}^{FCI}}{E_1}\times 100,
\end{align}
\begin{align}\label{CCC6}
	E_{2c}^{FCI}(\%E_2)=\frac{E_2-E_{2}^{FCI}}{E_2}\times 100,
\end{align}
where $E_{1c}^{FCI}(\%E_1)$ and $ E_{2c}^{FCI}(\%E_2)$ represent the percentual Coulomb contribution in the first and second excited states, respectively.
\begin{table}[H]
	\centering
	\begin{tabular}{|c|c|c|c|}
		\hline
		Aspect ratio & $E_{0c}^{FCI}(\%E_0)$ & $ E_{1c}^{FCI}(\%E_1)$ & $ E_{2c}^{FCI}(\%E_2)$ \\ \hline
		0.5 & 1.62 & 1.38 & 1.12 \\ \hline
		1 & 1.63 & 1.39 & 1.38 \\ \hline
		2 & 1.60 & 1.38 & 1.13 \\ \hline
	\end{tabular}
	\caption{Ground, first and second excited states of CdSe.}
	\label{Tablaa}
\end{table}
\begin{table}[H]
	\centering
	\begin{tabular}{|c|c|c|c|}
		\hline
		Aspect ratio & $E_{0c}^{FCI}(\%E_0)$ & $ E_{1c}^{FCI}(\%E_1)$ & $ E_{2c}^{FCI}(\%E_2)$ \\ \hline
		0.5 & 1.73 & 1.44& 1.24\\ \hline
		1 & 1.75 & 1.46 & 1.46\\ \hline
		2 & 1.71 & 1.44 & 1.20\\ \hline
	\end{tabular}
	\caption{Ground, first and second excited states of CdTe.}
\end{table}
\begin{table}[H]
	\centering
	\begin{tabular}{|c|c|c|c|}
		\hline
		Aspect ratio & $E_{0c}^{FCI}(\%E_0)$ & $ E_{1c}^{FCI}(\%E_1)$ & $ E_{2c}^{FCI}(\%E_2)$ \\ \hline
		0.5&1.39 & 1.27 & 0.84 \\ \hline
		1 & 1.40 & 1.26 & 1.26  \\ \hline
		2&1.37 & 1.27 & 0.98  \\ \hline
	\end{tabular}
	\caption{Ground, first and second excited states of ZnTe.}
	\label{Tablac}
\end{table}
In Tables (\ref{Tablaa}-\ref{Tablac}) and Figure (\ref{fig:Excited}), we can see the following characteristics:
\begin{itemize}
	\item The contributions of the Coulomb interaction to the eigenenergies decreases as the energy state increases.
	\item Contributions from Coulomb effects do not break the degeneracy in the energy levels.
	\item The first excited state obtained in the computational calculations is the lower part of the combination between the first and second excited state in the theoretical results shown in figure \ref{fig:enter-3}. This is lated to the change of shape of the quantum dots, which also explains why the curve of the second excited state in the computational calculations loses smoothness, under the influence of higher order excited states.
\end{itemize}
	
\section{Conclusions}
\label{sec:conclusiones}
Our study of spheroidal axially symmetric quantum dot reveals that self-energy effects (dielectric confinement) are neglected due to presence of one charge (electron-hole) counteracts the other (hole-electron). Furthermore, if this cancellation did not occur, the effects would be substantial, with values reaching $433.6 meV$, compared to the ground state energy  $\approx 1815 meV$ without taking in count the interactions between electron-hole.

The study of Coulomb correlations for the system has revealed that as the volume of confinement decreases, the electron-hole interactions become more pronounced, leading to higher eigenenergies, but increasing the basis set for configurations beyond a certain point does not significantly affect the results, indicating a diminishing contribution from additional configurations.
The results highlight the sensitivity of quantum dot energy levels to changes in shape and composition. It implies that even small alterations in the structure of the quantum dot can have a significant impact on the resulting energy states and excitations. 
Among the three studied types of QDs, it was found that the largest influence of the Coulomb interaction is in the CdTe's case.

\section{Acknowledgements}
The authors acknowledge financial support from  the Colombian SGR through project BPIN2021000191.
	
\section{Source Code Availability}
For transparency and reproducibility, we have made available the Python code used for the eigenvalue calculations on GitHub. Interested readers can find the repository at the following link: [https://github.com/CarlosABQ34/Energy-of-Qds.git]. The repository includes the main scripts and a README file with usage instructions.
	
\bibliographystyle{apsrev4-2}
\bibliography{referencias}  

		
		
		
		

\end{document}